\documentclass[sigconf]{acmart}

\usepackage[acronym,shortcuts]{glossaries}
\newacronym{FOV}{FOV}{field of view}
\newacronym{VPU}{VPU}{video processing unit}
\newacronym{VR}{VR}{virtual reality}
\newacronym{SPV}{SPV}{simulated prosthetic vision}
\newacronym{HMD}{HMD}{head-mounted display}

\AtBeginDocument{%
  \providecommand\BibTeX{{%
    \normalfont B\kern-0.5em{\scshape i\kern-0.25em b}\kern-0.8em\TeX}}}

\setcopyright{acmcopyright}
\copyrightyear{2021}
\acmYear{2021}
\acmDOI{10.1145/1122445.1122456}

\acmConference[AHs '21]{Augmented Humans '21}{February 22--24, 2021}{Online}
\acmBooktitle{Augmented Humans '21}
\acmPrice{15.00}
\acmISBN{978-1-4503-XXXX-X/18/06}

\begin{document}

\title{Towards Immersive Virtual Reality Simulations of Bionic Vision}

\author{Justin Kasowski}
\affiliation{
    \institution{University of California,}
    \city{Santa Barbara}
    \state{CA}
    \country{USA}
}
\email{justin_kasowski@ucsb.edu}

\author{Nathan Wu}
\affiliation{
    \institution{University of California,}
    \city{Santa Barbara}
    \state{CA}
    \country{USA}
}
\email{yangwu@ucsb.edu}

\author{Michael Beyeler}
\affiliation{
    \institution{University of California,}
    \city{Santa Barbara}
    \state{CA}
    \country{USA}
}
\email{mbeyeler@ucsb.edu}


\renewcommand{\shortauthors}{Kasowski, et al.}

\begin{abstract}
Bionic vision is a rapidly advancing field aimed at developing visual neuroprostheses (`bionic eyes') to restore useful vision to people who are blind.
However, a major outstanding challenge is predicting what people `see' when they use their devices.
% One key component of prosthetic vision is a limited field of view where the patient turns their head to update what the camera is "seeing".
The limited field of view of current devices necessitates head movements to scan the scene, which is difficult to simulate on a computer screen.
In addition, many computational models of bionic vision lack biological realism.
% However, many of the computational models used to develop the devices are oversimplified and lack both realism and biological accuracy.
To address these challenges, we propose to embed biologically realistic models of \acf{SPV} in 
immersive \acf{VR} so that sighted subjects can act as `virtual patients' in real-world tasks.
% \ac{VR} easily emulates this, but most \ac{SPV} models are monitor based.
% The current work embeds a biologically relevant model of prosthetic vision into \ac{VR} for a novel and realistic simulation. 
\end{abstract}




\begin{teaserfigure}
  \centering
  \includegraphics[width=0.8\textwidth]{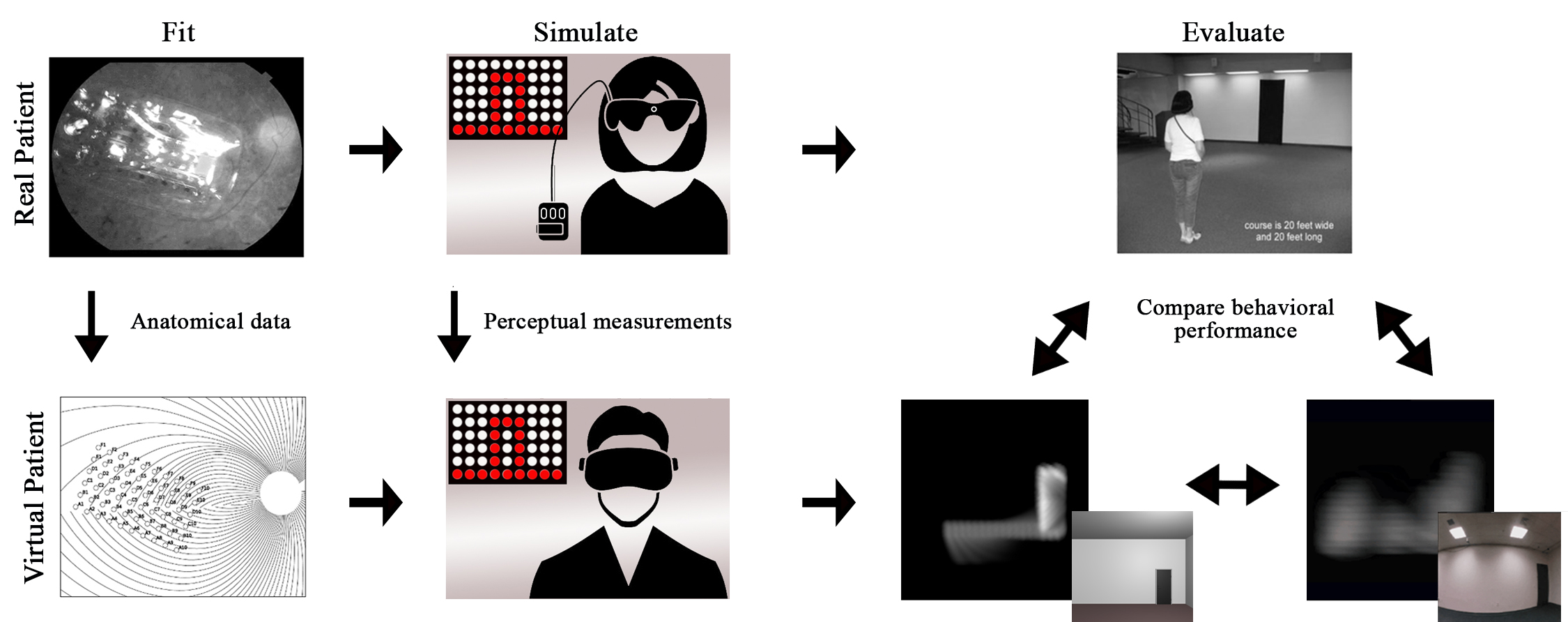}
  \caption{Virtual and real patients for bionic vision.
    \emph{Top row:} Retinal prosthesis patient. A microelectrode array is implanted in the eye to stimulate the retina \emph{(left)}. Light captured by an external camera is transformed into electrical pulses delivered to the retina to evoke visual percepts (%`phosphenes',
    \emph{middle}), which a patient uses to walk towards a door (\emph{right}).
    \emph{Bottom row:} Virtual patient. 
    Anatomical data is used to place a simulated implant on a simulated retina (\emph{left}). Visual input from a \acf{VR} device is used to generate realistic predictions of simulated prosthetic vision (SPV, \emph{middle}), which a virtual patient uses to walk to a simulated door in \acs{VR} \emph{(inner-right)}, or a door in the real world captured by the head-mounted display's camera  \emph{(outer-right)}. Edges stand out due to the specific preprocessing methods used, but a variety of methods can be tested. Behavioral performance can then be compared between real prosthesis patients, SPV of the real world, and SPV of the virtual world.
  }
  \Description{Virtual patients for bionic vision.}
  \label{fig:teaser}
\end{teaserfigure}

\maketitle

\section{Introduction}
Retinal degenerative diseases cause profound visual impairment in more than 10 million people worldwide, and retinal prostheses (`bionic eyes') are being developed to restore vision to these individuals. Analogous to a cochlear implant, these devices
convert video from a head-mounted camera into electrical pulses used to stimulate retinal neurons, which the brain interprets as visual percepts (`phosphenes'; Fig.~\ref{fig:teaser}, \emph{top row}).
% shortened..
% A head-mounted external camera feeds images to a \ac{VPU} where the signal is converted to electrode activation patterns and sent to an electrode array located in the eye. The signal from the \ac{VPU} is then used to electrically stimulate neurons and evoke visual percepts (‘phosphenes’).
Current devices have been shown to enable basic orientation \& mobility tasks \cite{ayton_update_2020}, but a growing body of evidence suggests that the vision restored by these devices differs substantially from normal sight
\cite{beyeler_learning_2017,erickson-davis_what_2020}.
% advances in the field could also boost performance in more complicated tasks. %Advances in neural interfaces have impact outside of bionic vision as well, showing promise for applications like complex robotics control, memory, and sense enhancement\cite{fernandez_development_2018}. 
% Despite the potential of these type of devices, there are still multiple limitations with restoring useful vision in daily life\cite{beyeler_learning_2017}

A major outstanding challenge is predicting what people `see' when they use their devices.
% In addition, since the external camera is fixed on the head, current retinal prostheses do not compensate for eye movements. % not sure we need these additional arguments
% Additionally, retinal prostheses are typically only implanted into one eye, resulting in a monocular visual field lacking depth perception. % that could make it easier to predict what they're seeing :)
Studies of \ac{SPV} often simplify phosphenes into small independent light sources  \cite{chen_simulating_2009, zapf_towards_2014, denis_simulated_2014} even though recent evidence suggests phosphenes vary drastically across subjects and electrodes \cite{beyeler_model_2019,erickson-davis_what_2020}.
% and often fail to assemble into complex  percepts \cite{erickson-davis_what_2020}. 
Another challenge is addressing the narrow \ac{FOV} found in most devices (but see \cite{ferlauto_design_2018}). This requires patients to scan the environment with head movements while trying to piece together the information \cite{erickson-davis_what_2020}, but many previous \ac{SPV} studies are performed on computer monitors. While some studies attempt to address this \cite{chen_simulating_2009, denis_simulated_2014, zapf_towards_2014}, most fail to account for phosphene distortions. It is therefore unclear how the findings of common \ac{SPV} studies would translate to real retinal prosthesis patients.

%2) no accurate phosphene model (what you said) => overhyped tech, conclusions of simulation papers don't generalize to real world
% An accurate model of prosthetic vision is invaluable as data collection with real patients is costly and time consuming. Previous models of prosthetic vision are largely based on the assumption that the eye performs as a pixel by pixel grid, a model commonly referred to as the “scoreboard model”. In this model, each electrode is represented by a single pixel with a fixed shape and size. In reality, phosphenes will have distortions in both space and time \cite{beyeler_learning_2017} and will vary with the frequency and amplitude of the electrical current \cite{nanduri_frequency_2012}. 

% The current work focuses on modeling these limitations by creating an immersive \ac{VR} simulation of bionic vision. This model takes into account multiple factors which have been previously ignored by other models. By using a simulation, normally-sighted volunteers can be turned into "virtual patients"; a process which will allow accurately studying all aspects of bionic vision.  

%\section{Related Work}

%If there's room, we should talk about how virtual patients have been used in other fields of medicine, or cite some of the limited VR work in bionic vision (draw from AR/VR review for references; Gislin Dagnelie comes to mind)

\section{Prototype}

To address these challenges, we embedded a biologically realistic model of \ac{SPV} \cite{beyeler_pulse2percept_2017,beyeler_model_2019} in immersive \ac{VR} using the Unity development platform, allowing sighted subjects to act as virtual patients in real-world tasks (see Fig.~\ref{fig:teaser}, \emph{bottom row}).
In this setup, the visual input about to be rendered to an HTC VIVE \ac{HMD} mimics the external camera of a retinal implant. This input can come from the \ac{HMD}'s camera or can be simulated in a virtual environment. A combination of compute and fragment shaders is used to simulate how this input is likely perceived by a real patient. Unlike previous models, our work is based on open-source code described in \cite{beyeler_learning_2017}, which generates a realistic prediction of \ac{SPV} that matches the field of view and distortions of real devices.
%Based on open-source code described in \cite{beyeler_pulse2percept_2017}, a combination of compute and fragment shaders was used to generate a realistic prediction of \ac{SPV} for each input frame. 
This allows sighted subjects to `see' through the eyes of a retinal prosthesis patient, taking into account their head and (in future work) eye movements as they explore an immersive virtual environment.
% Once validated, the model will be open-source and is designed to be flexible, such that 
Future \ac{SPV} models can be plugged in and applied to new prostheses once they become available.

\section{Research Goals}
% \subsection{Verify the model using simulated versions of visual prosthetic clinical trials}
\subsection{Provide realistic estimates of current bionic eye technologies}

% Second, current prostheses require patients to use head movements to scan their environment, a motif impossible to replicate with monitor-based simulations (e.g., \cite{kiral-kornek_improved_2014, zhao_recognition_2018, lu_estimation_2012}). Instead, immersive \ac{VR} environments allow us to focus on using \ac{SPV} for tasks that are known to diminish the quality of life for the blind (e.g. face recognition, navigation, reading, and self-care).

The prevailing approach to \ac{SPV} is to assume that activating a grid of electrodes leads to the percept of a grid of luminous dots, the brightness of which scales linearly with stimulus amplitude \cite{chen_simulating_2009,zapf_towards_2014,denis_simulated_2014}.
By ignoring percept distortions \cite{beyeler_model_2019,erickson-davis_what_2020}, performance predictions of such studies can be highly misleading.
In contrast, our work is constrained by neuroanatomical and psychophysical data.

In addition, current devices are typically evaluated on simple behavioral tasks, such as letter/object recognition \cite{cruz_argus_2013,edwards_assessment_2018}, following a line painted on the ground, or finding a door in an empty room \cite{humayun_preliminary_2009}.
Even simple letter recognition tasks require head movements to scan the scene, which is best emulated in an immersive \ac{VR} environment (note that \ac{FOV}\textsubscript{bionic eye} $<<$ \ac{FOV}\textsubscript{HMD}).
% by tailoring back 
% the significantly higher \acp{FOV} found in current \acp{HMD} to match those found in retinal prostheses.
% Behavioral performance of virtual patients can then be compared to real prosthesis patients.

% Tasks that are commonly used to evaluate device efficacy include , letter recognition \cite{cruz_argus_2013}, and object recognition \cite{edwards_assessment_2018}. Fig.~\ref{fig:teaser} shows the real life version of the door finding task\cite{humayun_preliminary_2009}, along with how it would appear under the current model of simulated prosthetic vision. Gathering data similar to that found in real trials will verify this model and allow for future testing with "virtual patients".  

% \subsection{Aim 2: Use virtual patients to further understand the current technology}
\subsection{Assess the potential of advanced stimulation strategies}
\begin{figure}[!t]
  \centering
  \includegraphics[width=\linewidth]{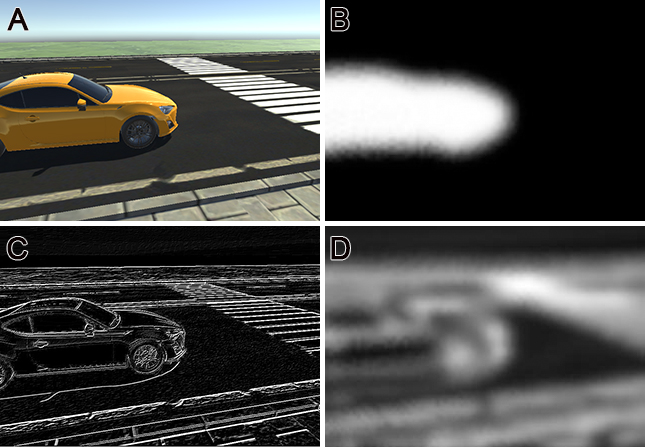}
  \caption{Different image processing techniques for \acf{SPV}. \emph{A)} Original Image. \emph{B)} \ac{SPV} after segmenting important objects. \emph{C)} Edge enhancement of the original image. \emph{D)} \ac{SPV} of the edge-enhanced image.}
  \label{fig2}
  \Description{Different preprocessing effects under simulated prosthetic vision}
\end{figure}

With the limited number of pixels found in current devices (e.g., Argus II: $6 \times 10$ electrodes), it is impossible to accurately represent a scene without preprocessing. Rather than aiming to restore ``natural vision'', there is potential merit in borrowing computer vision algorithms as preprocessing techniques to maximize the usefulness of bionic vision. Edge enhancement and contrast maximization are already routinely used in current retinal implants, and more advanced techniques based on object segmentation or visual saliency might further improve visual performance \cite{han_deep_2021}. 
% An accurate model of bionic vision allows for studying more advanced stimulation strategies, including a  plethora of software changes that could drastically change the patient's ability to interact with the world (Fig.~\ref{fig2}).
%Examples include facial recognition, text magnifiers, object segmentation, and other algorithms that may help people fully utilize their altered perception. 
For example, using an object detection algorithm for cars could help prostheses users at a crosswalk (Fig.~\ref{fig2}), while facial recognition or text magnification would be useful in other scenarios.

Although current bionic eyes have been implanted in over 500 patients worldwide, experimentation with improved stimulation protocols remains challenging and expensive. Virtual patients can offer an affordable alternative for designing high-throughput experiments that can test theoretical predictions, the best of which can then be validated in real prosthesis patients.

% Even with significant improvements in the technology, there will be a dramatic loss of information compared to non-prosthetic sight. Utilizing the available information is pivotal and can be readily studied with an accurate model of bionic vision.  

\subsection{Guide the prototyping of future devices}% using the knowledge gained from virtual patients}
The insights gained through virtual patients may help drive the changes for future devices. Obvious improvements could be realized by testing different electrode layouts and stimulation frequencies.
Some of these factors have been modeled before in isolation, but these models often predicted higher visual acuity than what was found in clinical trials \cite{chen_simulating_2009,cheng_advances_2017,zapf_towards_2014,denis_simulated_2014}. 
Therefore, using an established and psychophysically validated computational model of bionic vision may prove invaluable to generating realistic predictions of visual prosthetic performance.

% Other factors, like monocular implantation and eye movement, have largely been ignored as the vast majority of \ac{SPV} has been performed on a monitor. By utilizing \ac{VR}, these previously underassessed factors and their implications can be better understood. 

\section{Conclusion}

The present work constitutes a first essential step towards immersive \ac{VR} simulations of bionic vision.
% Our contribution is two-fold.
% First, in contrast to studies assuming that retinal implants lead to the perception of isolated, focal spots of light \cite{chen_simulating_2009,zapf_towards_2014,denis_simulated_2014}, we used a psychophysically validated computational model of bionic vision \cite{beyeler_pulse2percept_2017,beyeler_model_2019} to generate realistic predictions of \ac{SPV}.
% Second, current prostheses require patients to use head movements to scan their environment, a motif impossible to replicate with monitor-based simulations (e.g., \cite{kiral-kornek_improved_2014, zhao_recognition_2018, lu_estimation_2012}). Instead, immersive \ac{VR} environments allow us to focus on using \ac{SPV} for tasks that are known to diminish the quality of life for the blind (e.g. face recognition, navigation, reading, and self-care).
%
% By studying real-world tasks, the focus can shift from 'restoring natural vision', and rather focus on using the prosthetic vision efficiently in tasks that are known to diminish the quality of life for the blind (e.g. face recognition, navigation, reading, and self-care).
% Protocols found to benefit virtual patients can be validated with real patients and ultimately lead to an improved quality of life.
% Virtual patients can also be used to improve prosthetic design, including optimization of electrode configurations and stimulation protocols.
The proposed system has the potential to 1) further our understanding of the qualitative experience associated with different bionic eye technologies, 2) provide realistic expectations of bionic eye performance for patients, doctors, manufacturers, and regulatory bodies, and 3) accelerate the prototyping of new devices.

\bibliographystyle{ACM-Reference-Format}
\bibliography{references}

\end{document}